\documentclass[prb,twocolumn,amsmath,amssymb]{revtex4}
\usepackage{natbib}
\usepackage{graphicx}         
\usepackage{dcolumn}         
\usepackage{bm}                 

\begin{document}

\title{Spin-orbit coupling and phase-coherence in InAs nanowires}

\author{S. Est\'{e}vez Hern\'{a}ndez}
\affiliation{Institute of Bio- and Nanosystems (IBN-1) and
JARA-Fundamentals of Future Information Technology,
Forschungszentrum J\"ulich GmbH, 52425 J\"ulich, Germany}

\author{M. Akabori}
\altaffiliation[Present address: ] {Center for Nano-Materials and
Technology, Japan Advanced Institute of Science and Technology
(JAIST), 1-1, Asahidai, Nomi, Ishikawa 923-1292, Japan}
\affiliation{Institute of Bio- and Nanosystems (IBN-1) and
JARA-Fundamentals of Future Information Technology,
Forschungszentrum J\"ulich GmbH, 52425 J\"ulich, Germany}

\author{K. Sladek}
\affiliation{Institute of Bio- and Nanosystems (IBN-1) and
JARA-Fundamentals of Future Information Technology,
Forschungszentrum J\"ulich GmbH, 52425 J\"ulich, Germany}

\author{Ch. Volk}
\affiliation{Institute of Bio- and Nanosystems (IBN-1) and
JARA-Fundamentals of Future Information Technology,
Forschungszentrum J\"ulich GmbH, 52425 J\"ulich, Germany}

\author{S. Alagha}
\affiliation{Institute of Bio- and Nanosystems (IBN-1) and
JARA-Fundamentals of Future Information Technology,
Forschungszentrum J\"ulich GmbH, 52425 J\"ulich, Germany}

\author{H. Hardtdegen}
\affiliation{Institute of Bio- and Nanosystems (IBN-1) and
JARA-Fundamentals of Future Information Technology,
Forschungszentrum J\"ulich GmbH, 52425 J\"ulich, Germany}

\author{M. G. Pala}
\affiliation{IMEP-LAHC (UMR 5130), Grenoble INP, MINATEC, 3 Parvis
Louis N\'{e}el, BP 257, 38016 Grenoble, France}

\author{N. Demarina}
\altaffiliation[Permanent address: ] {Electronics Department,
Radiophysics Faculty, Nizhny Novgorod State University, Nizhny
Novgorod 603950, Russia} \affiliation{Institute of Bio- and
Nanosystems (IBN-1) and JARA-Fundamentals of Future Information
Technology, Forschungszentrum J\"ulich GmbH, 52425 J\"ulich,
Germany}

\author{D. Gr\"utzmacher}
\affiliation{Institute of Bio- and Nanosystems (IBN-1) and
JARA-Fundamentals of Future Information Technology,
Forschungszentrum J\"ulich GmbH, 52425 J\"ulich, Germany}

\author{Th. Sch\"apers}
\email{th.schaepers@fz-juelich.de} \affiliation{Institute of Bio-
and Nanosystems (IBN-1) and JARA-Fundamentals of Future
Information Technology, Forschungszentrum J\"ulich GmbH, 52425
J\"ulich, Germany}

\date{\today}

\hyphenation{InN}

\begin{abstract}
We investigated the magnetotransport of InAs nanowires grown by
selective area metal-organic vapor phase epitaxy. In the
temperature range between 0.5 and 30~K reproducible fluctuations
in the conductance upon variation of the magnetic field or the
back-gate voltage are observed, which are attributed to electron
interference effects in small disordered conductors. From the
correlation field of the magnetoconductance fluctuations the
phase-coherence length $l_\phi$ is determined. At the lowest
temperatures $l_\phi$ is found to be at least 300~nm, while for
temperatures exceeding 2~K a monotonous decrease of $l_\phi$ with
temperature is observed. A direct observation of the weak
antilocalization effect indicating the presence of spin-orbit
coupling is masked by the strong magnetoconductance fluctuations.
However, by averaging the magnetoconductance over a range of gate
voltages a clear peak in the magnetoconductance due to the weak
antilocalization effect was resolved. By comparison of the
experimental data to simulations based on a recursive
two-dimensional Green's function approach a spin-orbit scattering
length of approximately 70~nm was extracted, indicating the
presence of strong spin-orbit coupling.
\end{abstract}

\maketitle

\section{Introduction}

Semiconductor nanowires fabricated by a bottom-up
approach\cite{Thelander06,Lu06,Ikejiri07} offer an exciting route
to study fundamental quantum effects in electron transport of
nanostructures.\cite{DeFranceschi03,Fasth05a,Pfund06,VanDam06,Fuhrer07,Richter08}
By choosing a narrow-band gap semiconductor for the nanowire, i.e.
InAs or InN, the common problem connected with carrier depletion
at the surface can be
avoided.\cite{Thelander06,Chang05,Calarco07,Bloemers08,Ford09} As
illustrated in Fig.~\ref{fig:1}(a), for these semiconductors a
surface accumulation due to the Fermi-level pinning in the
conduction band occurs.\cite{Lueth10} Here, due to the electron
transfer from the surface states the conduction and valence bands
are bent downwards and a two-dimensional electron gas (2DEG) is
formed at the surface. The electrons are confined in a
triangularly-shaped potential well. In the case of InAs-based
semiconductors it is known that the electrons in the accumulation
layer are also subject to the impact of a strong spin-orbit
coupling.\cite{Nitta97,Engels97,Schierholz04} The macroscopic
electric field in the triangularly-shaped quantum well lifts the
spin-degeneracy for propagating electrons. This so-called Rashba
effect\cite{Bychkov84} is of relevance for spin electronic
devices, since it can be employed for gate-controlled spin
manipulation.\cite{Datta90} It was shown theoretically that by
making use of quantum wire structures the performance of spin
transistor structures with respect output modulation can be
improved and devices with new functionalities, e.g. spin filters,
can be realized.\cite{Datta90,Bournel98,Governale02b,Zuelicke02}
Experimentally the presence of spin-orbit coupling can be verified
by the observance of the weak antilocalization
effect.\cite{Bergmann84,Hikami80} For one-dimensional structures
the spin-orbit scattering length $l_{so}$ being a measure of the
strength of spin-orbit coupling can be determined by fitting an
appropriate theoretical model.\cite{Kurdak92,Kettemann07,Wenk10}

At low temperatures the electron transport in a nanowire is
governed by electron interference. The relevant parameter
connected to this phenomenon is the phase-coherence length
$l_\phi$. If the length of the nanowire is comparable to $l_\phi$
one finds conductance fluctuates with an amplitude on the order of
$e^2/h$ if a magnetic field is applied or if the electron
concentration is changed by means of a gate
electrode.\cite{Hansen05,Bloemers08,Bloemers08a} The fluctuations
originate from the fact that in small disordered samples, electron
interference effects are not averaged
out.\cite{Altshuler85b,Lee87} Although fluctuations in the
conductance can mask other quantum effects, i.e. weak localization
or quantized conductance, important information on the electron
transport can be gained from the statistical properties of the
fluctuations themselves.

We studied the transport properties of small individual InAs
nanowires and demonstrate that in spite of the fact that the
low-temperature magneto-conductance strongly fluctuates the
essential transport parameters, e.g. $l_\phi$ or the spin-orbit
scattering length $l_{so}$ can be extracted. Information on
$l_\phi$ at various temperatures is obtained from the average
fluctuation amplitude as well as from the correlation field. As
mentioned above, for electrons confined at the surface a strong
Rashba spin-orbit coupling is expected, due to the high electric
field in the quantum well.\cite{Engels97} Previously, for
individual InAs wires\cite{Dhara09,Roulleau10} or InAs wires
contacted in parallel\cite{Hansen05} a maximum of the
magnetoconductance at zero field due to weak antilocalization was
reported indicating the presence of spin-orbit coupling, whereas
no weak antilocalization in InAs nanowires was found by
others.\cite{Liang09}

It is known for InAs nanowires that depending on the growth
conditions different crystal structures form, which have a direct
impact on the transport properties.\cite{Schroer10} Our nanowires
were grown by using selective-area metal-organic vapor phase
epitaxy. In this case no metallic particles were used in contrast
to the vapor-liquid-solid growth which is mostly employed for InAs
nanowire growth.\cite{Thelander06} Since our method is different
we were interested if indications spin-orbit coupling can be
observed as well. However, the growth method we used limits
maximum nanowire length to a few micrometers.\cite{Akabori09} The
short nanowire length is the reason that strong conductance
fluctuations occur in our magnetotransport experiments masking the
weak antilocalization effect. Nonetheless, by making use of
averaging the conductance with respect to the gate
voltage,\cite{Petersen09} we were able to resolve weak
antilocalization. Thus by the method applied here one can assess
the strength of spin-orbit coupling even in the presence of
conductance fluctuations. The measured characteristics of the weak
antilocalization is compared to calculated characteristics based
on a recursive real-space Green's function approach. In contrast
to previous theoretical models treating quantum wires based on a
planar 2DEG, here the cylindrical symmetry of the surface 2DEG is
taken explicitly into account.

We organized our article as follows: In Sect.~\ref{Sect:Exp}
details on nanowire growth and the sample preparation are given.
Simulations of quantum levels at the InAs interface are presented
in Sect.~\ref{Sect:Sim}, while in Sect.~\ref{Sect:Res} the
experimental results are discussed. Concluding remarks are given
in Sect.~\ref{Sect:Con}.

\section{Experimental} \label{Sect:Exp}

\begin{figure}[]
\begin{center}
\includegraphics[width=0.8\columnwidth]{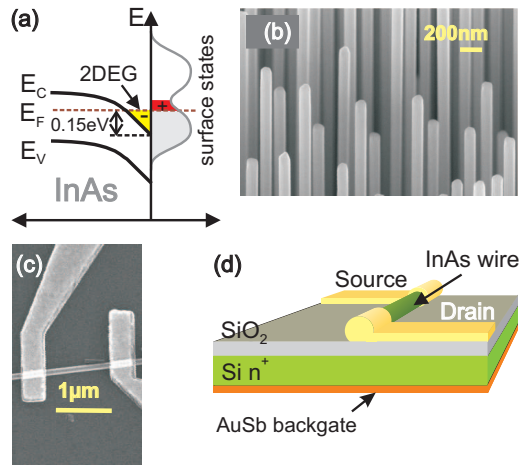}
\caption{(color online) (a) Illustration of the formation of a
two-dimensional electron gas (2DEG) at the interface. The
electrons are transferred from the surface states. The conduction
and valence band $E_c$ and $E_v$, respective are bent downwards to
form a triangular-shaped quantum well. (b) Scanning electron beam
micrograph of the as-grown InAs nanowires (c) Scanning electron
beam micrograph of the contacted nanowire. (d) Schematic
illustration of the sample. } \label{fig:1}
\end{center}
\end{figure}

The InAs nanowires were selectively grown on a patterned GaAs
(111)B substrate by low-pressure metal-organic vapor phase epitaxy
in an N$_2$ atmosphere at a temperature of
650$^\circ$C.\cite{Akabori09} The wires have a diameter $d$ of
approximately 100~nm. Figure~\ref{fig:1}(b) shows a scanning
electron micrograph of the as-grown InAs nanowires.

For the transport measurements contact pads and adjustment markers
were defined on an $n$-type doped Si (100) wafer covered by a
100-nm-thick SiO$_2$ layer. The SiO$_2$ layer was prepared by
thermal oxidation. The InAs nanowires were dispersed on the
patterned substrate, by separating them first from their host
substrate in an acetone solution. Subsequently, a droplet of
acetone containing the detached InAs nanowires was put on the
patterned substrate. The nanowires were contacted individually by
a pair of Ti/Au electrodes using electron beam lithography. In
order to improve the contact resistance Ar$^+$ sputtering was
employed prior to the metal deposition. The substrate was used as
a back-gate to control the electron concentration. The contact
separation $L$ of the wire reported here was 1~$\mu$m. In
Fig.~\ref{fig:1}(c), an electron beam micrograph of the sample is
shown, while in Fig.~\ref{fig:1}(d) a schematics of the sample
layout can be found. In total three nanowires were measured which
showed similar properties as the one presented here. At zero gate
voltage the nanowire has a resistance of 30.0~k$\Omega$ after
subtracting the typical contact resistance of 1~k$\Omega$. The
contact resistance and resistivity $\rho$ was estimated by
resistance measurements at 4~K of many nanowires of the same
growth run with different contact separations. The typical
electron concentration $n_{3D}$ of $2.4 \times 10^{17}$~cm$^{-3}$
was determined by extracting the threshold voltage from the
low-temperature transfer characteristics on more than ten samples
using the method described in Ref.~[\onlinecite{Dayeh07}]. For the
mobility $\mu=1/(en_{3D}\rho)$ we obtained a low-temperature value
of 3300~cm$^2$/Vs. The value of the diffusion constant
$\mathcal{D}=v_F^2 \tau/3$ , with the Fermi velocity $v_F$
determined from $n_{3D}$ and the elastic scattering time $\tau$
calculated from $\mu$, is equal to $1.3 \times 10^{-6}$~cm$^2$/s.
The Fermi energy $E_F$ is estimated to be about 60~meV.

The magneto-transport measurements were performed in a He-3
cryostat at temperatures ranging from 0.5~K to 30~K. The
magnetoresistance was measured by using a lock-in technique with
an ac bias current of 5~nA and a magnetic field oriented
perpendicular to the wire axis.

\section{Simulation} \label{Sect:Sim}

\begin{figure}[]
\begin{center}
\includegraphics[width=1.0\columnwidth]{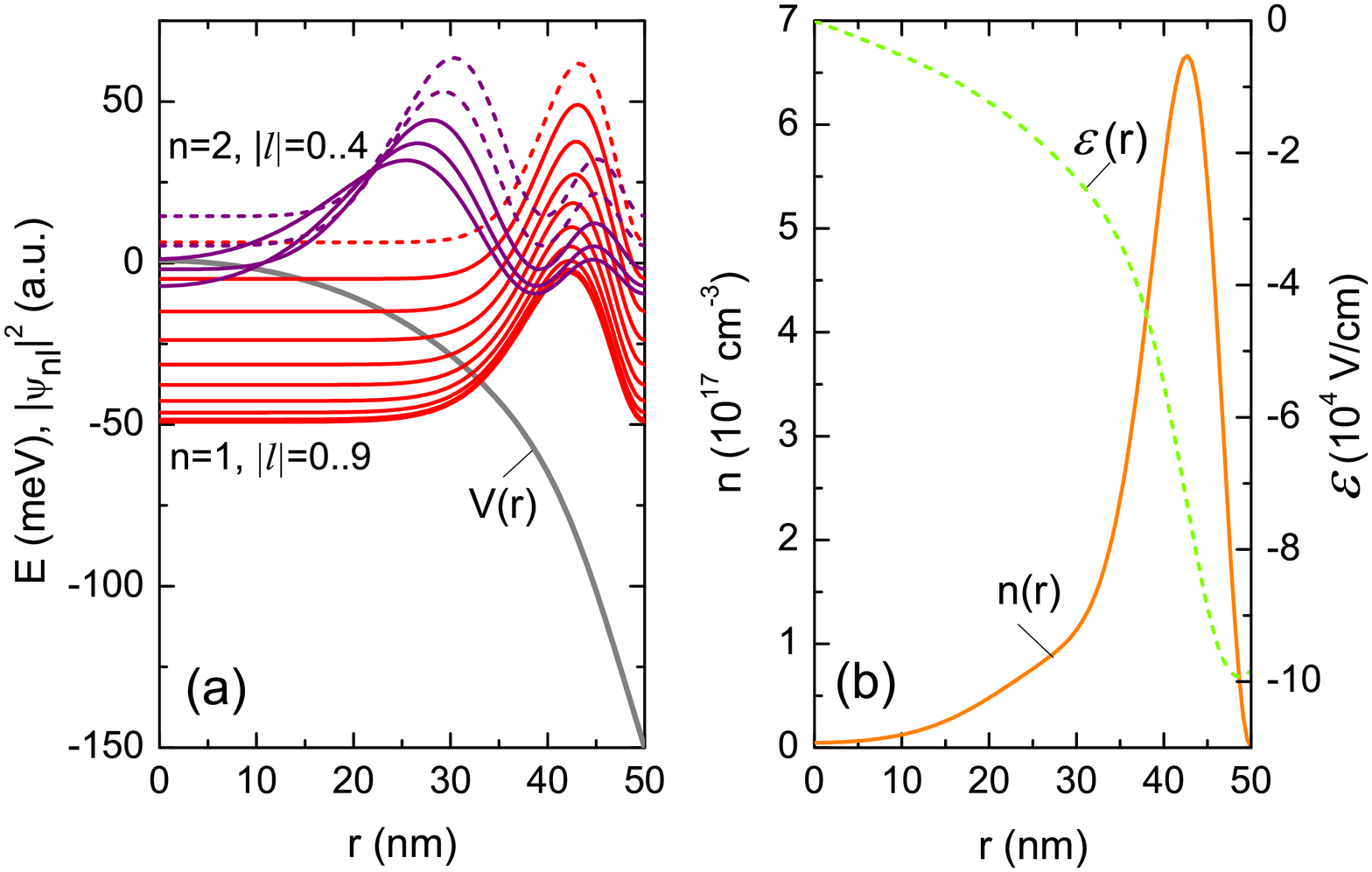}
\caption{(Color online) (a) Calculated profile of the conduction
band $V(r)$ and probability density $|\psi_{nl}|^2$ for $n=1$, 2
at various angular momentum numbers $l$. The intersection of
$|\psi_{nl}|^2$ at $r=0$ reflects the energy of the eigenstates.
(b) Total electron density $n(r)$ and electric field $\mathcal{E}$
as a function of radius.} \label{fig:2}
\end{center}
\end{figure}

In order to illustrate the interface accumulation layer of an InAs
nanowire with Fermi level pinning within the conduction band, we
calculated the quantum states in a nominally undoped InAs nanowire
at $B=0$. The calculations were performed by employing a
Schr\"odinger--Poisson solver dedicated to cylindrically shaped
conductors. The contribution of disorder, i.e. due to impurities
or interface roughness was not taken into account. For the surface
Fermi level pinning a value of 0.15~eV was assumed, which is a
common value for InAs.\cite{Smit89,Lueth10} At the interface an
infinite potential barrier was taken. In order to account for the
conduction band non-parabolicity, we inserted an enhanced
effective mass of $0.033\; m_0$. Instead of employing an
energy-dependent effective mass, we used a simpler approach
proposed by Schierholz \emph{et al.},\cite{Schierholz04} where the
average effective mass is a function of the electron
concentration. Using these parameters the calculated integrated
electron density agreed well with the experimental average
concentration of $2.4 \times 10^{17}$~cm$^{-3}$. The calculated
conduction band profile $V(r)$ is shown in Fig.~\ref{fig:2}(a)
together with the probability density $|\psi_{nl}|^2$ for the
first ($n=1$) and second ($n=2$) subbands at different angular
momenta $l=0,\pm 1, \pm 2, \pm 3,\ldots$. Due to the Fermi level
pinning at the surface, a potential well with an approximately
triangular shape is formed. This is also reflected by the electron
density $n(r)$ shown in Fig.~\ref{fig:2}(b), where a surface
accumulation layer is formed with a maximum value of $n(r)$ being
8~nm below the surface. As can also be seen in
Fig.~\ref{fig:2}(b), the absolute value of the electric field
increases monotonously for increasing radius, reaching a value of
about $-7.3 \times 10^4$~V/cm where $n(r)$ reaches its maximum.
\begin{figure}[]
\begin{center}
\includegraphics[width=1.0\columnwidth]{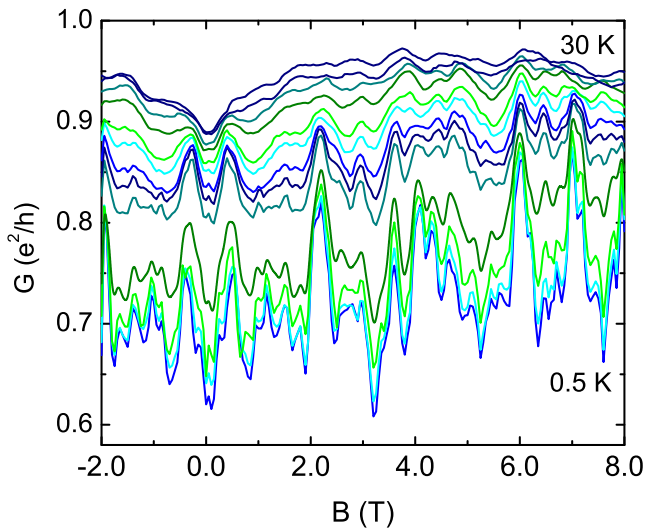}
\caption{(color online) (a) Fluctuating magneto-conductance in
units of $e^2/h$ at temperatures of 0.5, 1, 2, 3, 4, 5, 6, 8, 10,
15, 20, 25, and 30~K, at zero gate voltage.} \label{fig:3}
\end{center}
\end{figure}

\section{Experimental Results and Discussion} \label{Sect:Res}

We will first focus on the transport properties at zero gate bias,
in order to obtain information on the phase-coherence length
$l_\phi$. In Fig.~\ref{fig:3} the magneto-conductance is plotted
for various temperatures ranging from 0.5 to 30~K. 
\begin{figure}[]
\begin{center}
\includegraphics[width=0.8\columnwidth]{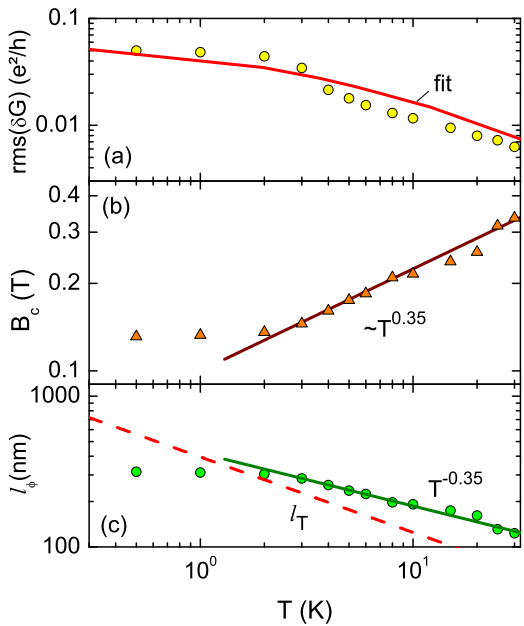}
\caption{(Color online) (a) root-mean-square $\mathrm{rms}(\delta
G)_B$ of the magneto-conductance as a function of temperature
(circles). The solid line represents the fit following
Ref.~[\onlinecite{Beenakker88a}] using the values of $l_\phi$
obtained from $B_c$. (b) Correlation field $B_c$ vs. temperature
$T$ (triangles). The solid line represents the increase of
proportional to $T^{0.35}$. (c) Phase-coherence length $l_\phi$ determined from the
correlation field $B_c$ (dots). The solid line represents the
decrease of $l_\phi$ proportional to $T^{-0.35}$. The broken line
corresponds to the thermal length $l_T$.} \label{fig:4}
\end{center}
\end{figure}
As one can see,
on average the conductance increases with temperature. At low
temperature the conductance strongly fluctuates, while at higher
temperatures these fluctuations are largely smeared out. The
average fluctuation amplitude around the mean value of the
conductance is quantified by the root-mean-square ${\rm
rms}(\delta G)_{B}=\sqrt{{\rm var} (\delta G)}_{B}$. Here, the
variance is defined by ${\rm var}(\delta G)_{B}=\langle \delta
G^2\rangle_{B}$, with $\langle \cdots \rangle_{B}$ the average of
the conductance fluctuations $\delta G$ over $B$. As can be seen
in Fig.~\ref{fig:4}(a), at temperatures below about 2~K the
average amplitude is saturated, while above that temperature ${\rm
rms}(G)_B$ decreases monotonously. The corresponding correlation
field $B_c$ of the fluctuations is shown in Fig.~\ref{fig:4}(b).
The correlation field was extracted from the autocorrelation
function of $\delta G$ given by $F(\Delta B)=\langle \delta
G(B+\Delta B)\delta G (B)\rangle $, with $B_c$ corresponding to
half maximum of the autocorrelation function $F(B_c)=\frac{1}{2}
F(0)$.\cite{Lee87} At low temperatures ($T\leq2$~K) we find that
$B_c$ is approximately constant at 0.13~T, while it increases
proportional to $T^{0.35}$ for temperatures larger than 2~K. Under
the assumption that $l_\phi$ is larger than the diameter $d$ of
the nanowire one can extract $l_\phi$ directly from correlation
field using the expression\cite{Beenakker88a}
\begin{equation} l_\phi=\gamma
(h/e)(1/B_cd)\; \label{eq:1} ,
\end{equation}
where $\gamma$ is a pre-factor depending on the transport regime.
The area $l_\phi d$ corresponds to the maximum area enclosed by
phase-coherent time-reversed electron waves penetrated by a single
flux quantum $h/e$. Here, we assumed the dirty metal limit with
$\gamma=0.95$.\cite{Beenakker88a} At temperatures below about 2~K,
$l_\phi$ has its maximum value at about 300~nm, while at higher
temperatures $l_\phi$ decreases proportional to $T^{-0.35}$ to a
value of slightly above 100~nm at $T=30$~K. The cylindrical
geometry of our nanowire does not correspond completely to the
one-dimensional conductors the above employed theory is based on.
Therefore, the parameter $\gamma$ and thus $\l_\phi$ might differ.
In fact, as one can infer from Fig.~\ref{fig:4}(b) up to 2~K the
correlation field $B_c$ is constant. A possible reason for this
might be, that phase-coherent transport is rather limited by the
contact separation.\cite{Bloemers08} Therefore the actual values
of $l_\phi$ are presumably larger than the ones calculated from
Eq.~(\ref{eq:1}).

The values of $l_\phi$ gained from the analysis of $B_c$ were
taken to calculate the theoretically expected values for
$\mathrm{rms}(\delta G)$. For this purpose we used the
interpolation formula of Beenakker and van Houten given
by:\cite{Beenakker88a}
\begin{equation}
\mathrm{rms}(\delta G)=\beta \frac{e^2}{h} \left( \frac{l_\phi}{L}
\right)^{3/2}
\left[1+\frac{9}{2\pi}\left(\frac{l_\phi}{l_T}\right)^2\right]^{-1/2}
\; . \label{eq:2}
\end{equation}
Here, $l_T=\sqrt{\hbar \mathcal{D}/k_BT}$ is the thermal length.
Theoretically, for $\beta$ a value of $\sqrt{3}$ is expected for
the situation that a magnetic field is applied and spin-orbit
coupling is present. As can be seen in Fig.~\ref{fig:4}(a), a
satisfactory agreement of the change with temperature between
theory and the experimental values of $\mathrm{rms}(\delta G)_B$
can be obtained, however for $\beta$ a value of 0.34 had to be
inserted being a factor of 5 smaller than the theoretically
expected value. A possible reason for this discrepancy might also
be that in our case the electrons propagate through in a conductor
of cylindrical shape, while the theory was developed for a quantum
wire based on a two-dimensional electron gas.

\begin{figure}[]
\begin{center}
\includegraphics[width=1.0\columnwidth]{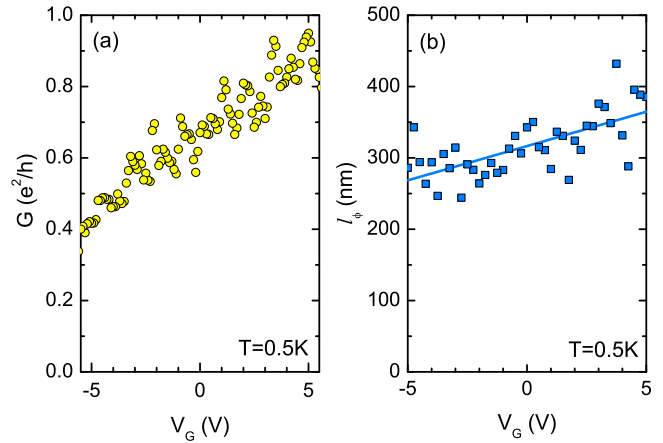}
\caption{(Color online) (a) Normalized conductance as a function
of gate voltage at zero magnetic field. The measurement
temperature was 0.5~K. (b) Phase-coherence length $l_\phi$
extracted from the correlation field as a function of gate
voltage.} \label{fig:5}
\end{center}
\end{figure}

\begin{figure}[]
\begin{center}
\includegraphics[width=0.85\columnwidth]{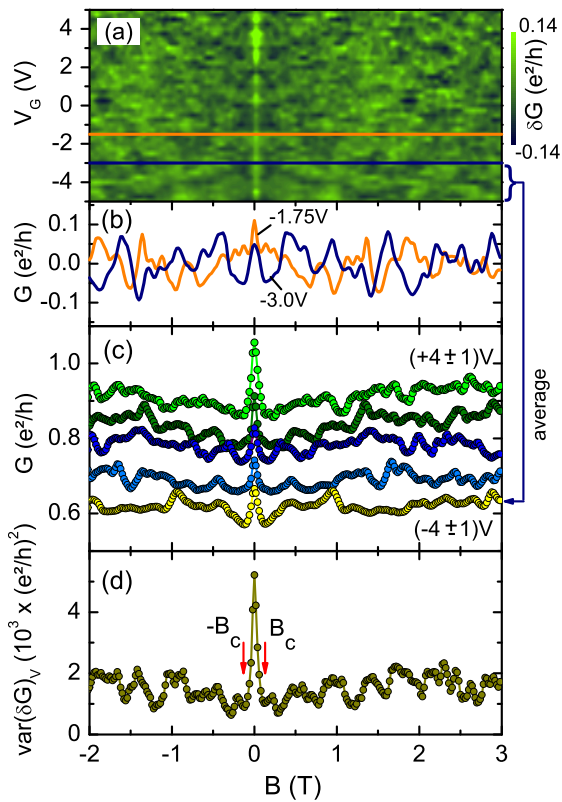}
\caption{(Color online) (a) Color-scaled conductance as a function
of magnetic field and gate voltage at 0.5~K. The trace of the
conductance fluctuations shown in (b) are indicated by the
horizontal lines. (a) Magneto-conductance at 0.5~K at
$V_G=-1.75$~V (red curve) and at $-3.0$~V (blue curve). (c)
Averaged magneto-conductance corrections for fluctuation
measurements within a 2~V-wide gate voltage interval. From top to
bottom the center-gate-voltages of the intervals are: $+4, +2, 0,
-2, -4$~V. The range used for the $-4$~V center-gate-voltage is
indicated by the arrow. (d) Variance $\mathrm{var}(G)_{V}$ as a
function of the magnetic field at 0.5~K. The arrows indicate the
position of the correlation field $B_c$.} \label{fig:6}
\end{center}
\end{figure}

By applying a gate voltage the electron concentration and thus the
conductance can be adjusted. As can be seen in
Fig.~\ref{fig:5}(a), by varying $V_G$ from $-5$~V to $+5$~V the
conductance $G$ increases from $0.4~e^2/h$ to almost $1.0~e^2/h$.
On average $G$ increases linearly with $V_G$. The weak dependence
of the conductance on the gate voltage can be related to the large
thickness of the SiO$_2$ layer and to a large density of interface
states. The relatively large variations of $G$ can be attributed
to conductance fluctuations originating from changes in the
electron interference due to the variation of the Fermi wave
length.\cite{Lee87} At fixed gate voltages the conductance
fluctuations were recorded as a function magnetic field [cf.
Fig.~\ref{fig:6}(a) and (b)]. One can clearly see, that owing to
the two-terminal measurement set-up the fluctuations are symmetric
upon reversal of the magnetic field. At different fixed gate
voltages $V_G$ the phase-coherence length $l_\phi$ was determined
from $B_c$ of the magneto-conductance fluctuations curves. As can
be seen in Fig.~\ref{fig:5}(b), the resulting values of $l_\phi$
increase with increasing $V_G$ and thus increasing electron
concentration. This increase can be attributed to the decrease of
electron-electron scattering rate $1/\tau_{ee}$ with increasing
Fermi energy.\cite{Altshuler79}

The measurements of the conductance fluctuations as a function of
magnetic field were used directly to obtain information on the
phase-coherence length $l_\phi$, however no phenomena related to
spin-effects, i.e. a distinct peak at $B=0$ due to weak
antilocalization, are resolved. This is illustrated in
Fig.~\ref{fig:6}(b), where the magnetoconductance fluctuations
$\delta G(B)$ at a temperature of 0.5~K are plotted for two
different back-gate voltages. In both curves no clear peak at zero
magnetic field exceeding the average fluctuation amplitude is
observed as it is expected for the weak antilocalization effect.
In order to resolve weak antilocalization, we made use of
averaging uncorrelated fluctuation patterns at different back-gate
voltages and thus suppressing the fluctuation amplitude
significantly. The underlying data set is shown in
Fig.~\ref{fig:6}(a). The conductance fluctuation patterns are
expected to be completely uncorrelated if the change of the Fermi
energy $E_F$ induced by the gate voltage exceeds the correlation
energy $E_c$. For the case $l_\phi < L$ the correlation energy is
given by $E_c=h \mathcal{D}/l_\phi^2$.\cite{Lee87} By taking the
lower bound value of $l_\phi \approx 300$~nm at low temperatures
one obtains $E_c$ of about 0.6~meV. Taking to account the value of
$E_F$ this corresponds to a gate voltage interval $V_c$ of 0.1~V
in which the transport is correlated.

In Fig.~\ref{fig:6}(c) the magneto-conductance curves from the
averaging in successive gate voltage intervals with a width of 2~V
are shown. The averaging was performed on magneto-conductance
traces for gate voltages each differing by 0.25~V being larger
than $V_c$. A comparison with the single magneto-conductance
curves at fixed gate voltages [c.f. Fig.~\ref{fig:6}(b)] confirms
that a considerable damping of the fluctuation amplitude is
achieved. Since the gate voltage difference of 0.25~V is larger
than $V_c$ estimated above, it is assured that the averaging is
performed over statistically independent traces. The choice of the
gate voltage interval was a trade-off between an effective
averaging to suppress the conductance fluctuations and a
sufficiently small change of the phase-coherence length. The
chosen width of 2~V meet this requirement. As can clearly be seen
in Fig.~\ref{fig:6}(c), all five traces reveal a clear peak at
$B=0$, which can be attributed to the weak antilocalization
effect. When comparing the different gate voltage ranges we find,
that for larger gate voltage and thus larger electron
concentrations the peak at $B=0$ is higher.

\begin{figure}[h!]
\begin{center}
\includegraphics[width=1.0\columnwidth]{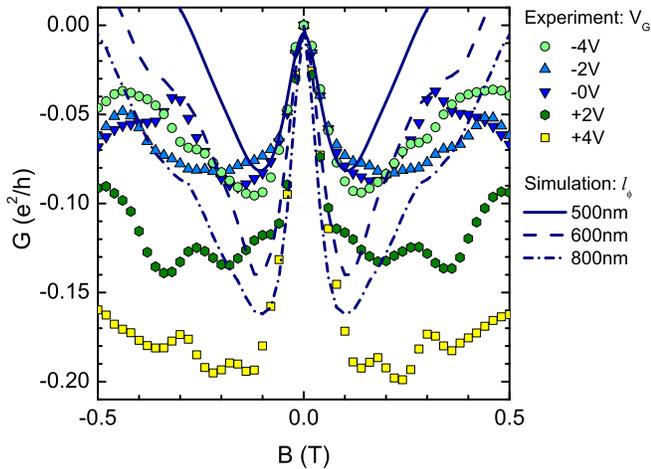}
\caption{(Color online) Averaged magneto-conductance corrections
for fluctuation measurements within a 2~V-wide gate voltage
interval (dots). The center-gate-voltages of the intervals are:
$+4, +2, 0, -2, -4$~V. The lines represent the corresponding
calculations based on the recursive Green's function approach.}
\label{fig:7}
\end{center}
\end{figure}

The averaged magneto-conductance curves can be used to estimate
the spin-orbit scattering length $l_{so}$, being a measure for the
strength of spin-orbit coupling. In Fig.~\ref{fig:7} the averaged
magneto-conductance correction is plotted for the five different
gate voltage intervals together with the results of numerical
simulations based on a recursive two-dimensional real-space
Green's function approach,\cite{Schaepers06} where the Rashba
spin-orbit coupling is modelled via an effective Hamiltonian
$H_R=\alpha_R \, \vec{\sigma} \times \vec{k}$. Here, $\alpha_R$ is
the Rashba coupling parameter, being a measure of the strength of
spin-orbit coupling and associated to the spin-orbit length
$l_{so}=\hbar^2/(2m^*\alpha_R)$, $\sigma$ are the Pauli spin
matrices, and $\vec{k}$ is the electron wave vector. In order to
take into account the cylindrical geometry of the wires, periodic
boundary conditions are imposed to mimic a plane wave propagation
along the transverse direction, as well as a sinusoidal external
magnetic field. In our zero-temperature calculations we assume the
presence of purely elastic scattering and the wire length $L$
plays the role of dephasing length $l_\phi$.\cite{Schaepers06} The
numerical scheme consists in dividing the system into
one-dimensional slices along the transport direction and to
compute the isolated Green's functions of each slice. Hence, the
total Green's function $G^R$ is obtained by recursively connecting
the on-site Green's function via a procedure based on the Dyson
equation. Ohmic contacts are included by means of self-energies
$\Sigma_{L/R}$ obtained from the surface Green's function of
semi-infinite leads. We assume that the external leads attached to
our system are subject to the same homogeneous magnetic field as
is present in the sample, but have no spin-orbit coupling.
Magnetic field in the channel is accounted for by multiplying
Peierls phases to the hopping elements of the
Hamiltonian.\cite{Baranger91} The conductance is expressed by the
Landauer-B\"uttiker formula $G=e^2/h \sum_{n,n'}\sum_{n,n'} T_{n'
n}^{s' s}$ where the sum runs over all incoming and outgoing
channels, and $T_{n' n}^{s' s}$ is the transmission probability
from mode $n$ with spin $s$ to mode $n'$ with spin $s'$.
Backscattering is induced by an effective potential due to
randomly distributed obstacles whose density is tuned, in order to
fit the experimental mean free path. Hence, magnetoconductance
corrections in Fig.~\ref{fig:7} are obtained by averaging hundreds
of simulations with different obstacle realizations for an energy
range close to the Fermi level.

As can be seen in Fig.~\ref{fig:7} a satisfactory agreement
between the experimentally observed weak-antilocalization peak
width and the simulation was achieved for a spin-orbit scattering
length of $l_{so}=70$~nm. In our simulation we systematically
increased $l_\phi$ from 500~nm, where the resulting peak height
fits to the experimental curves at $V_G=-4$~V, $-2$~V, and 0, to a
maximum values of $l_\phi=800$~nm. The experimental curve
belonging to $V_G=+2$~V fits reasonably well to the simulated
curve for $l_\phi=600$~nm, while the peak experimental curve for
$+4$~V is even larger that the calculated curve for
$l_\phi=800$~nm. The steeper decrease of the simulations beyond
the conductance minimum is due to the choice of the obstacle
potential whose gaussian shape induces a high-field localization
stronger than in the real wire. We notice that in our simulations
we have compared the experimental data by adopting a
phase-coherence length systematically larger than the experimental
values plotted in Fig.~\ref{fig:4}(c). As argued above the values
of $l_\phi$ obtained by employing Eq.~(\ref{eq:1}) might be too
small. On the other hand, the discrepancy can also be due to some
geometric effect which is not properly described in our model. As
can be inferred from the comparison of the experimental data and
the simulations Fig.~\ref{fig:7}, the resulting values of $l_\phi$
increase with increasing $V_G$ and thus increasing electron
concentration. The increase of $l_\phi$ with increasing electron
concentration can be attributed to the decrease of
electron-electron scattering rate $1/\tau_{ee}$ with increasing
Fermi energy.\cite{Altshuler79}

The conclusion that spin-orbit scattering is present in our wires
is also supported by the analysis of the variance
$\mathrm{var}(\delta G)_V$ with respect to the gate voltage taken
at different magnetic fields. As can be seen in
Fig.~\ref{fig:6}(d), $\mathrm{var}(\delta G)_V$ vs. $B$ shows only
a single peak at $B=0$. This is expected from theory for systems
comprising a strong spin-orbit coupling.\cite{Altshuler86,Stone89}
In this regime the appearance of universal conductance
fluctuations is due to the interference of correlated electron
pairs in the singlet state only, while the triplet state is
suppressed due to spin-orbit coupling. The singlet state can be
subdivided in the so-called cooperon and diffusion channels.
Applying a magnetic field results in a complete suppression of the
magnetic flux sensitive cooperon channel. In fact, the
experimentally observed drop of $\mathrm{var}(\delta G)_V$ by
about a factor of 2.5 is close to the theoretically expected drop
by 2. As indicated in Fig.~\ref{fig:6}(d) the half-width of the
peak corresponds to the value of the correlation field
$B_c=0.13$~T. This feature is in accordance with
theory.\cite{Altshuler86,Stone89} Only in the case of systems with
weak spin-orbit coupling, a second distinct decrease of the
variance originating from the influence of the Zeeman effect at a
characteristic field $B_{c2}=E_c/g\mu_B$ is expected. Here, $g$ is
the gyromagnetic factor and $\mu_B$ the Bohr magneton. By assuming
$g=14.5$ for InAs and $E_c=0.6$~meV, as has been estimated above,
one obtains a value of 0.7~T for $B_{c2}$. Obviously, by
inspecting $\mathrm{var}(\delta G)_V$ in Fig.~\ref{fig:6}(d) one
does not find any indication of this second drop. This strongly
supports our assumption of the presence of strong spin-orbit
coupling in InAs nanowires.

From magnetotransport studies on two-dimensional electron gases
formed in the surface accumulation layer of InAs it is well-known
that the strong Rashba effect results in a short spin-orbit
scattering length $l_{so}$. By investigating the characteristic
beating pattern in the Shubnikov--de Haas
measurements\cite{Matsuyama00} or by measuring the weak
antilocalization effect \cite{Schierholz04} values of $l_{so}$ in
the order of 100~nm were found being also confirmed by theoretical
calculations.\cite{Lamari01} In InAs nanowires weak
antilocalization was also observed by Hansen \emph{et
al.},\cite{Hansen05}, here the suppression of conductance
fluctuations was achieved by measuring many nanowires in parallel.
Typical values of $l_{so}$ in the order of 200~nm were extracted.
On longer InAs nanowires where the conductance fluctuations are
sufficiently suppressed $l_{so}$ values between 200 and 100~nm
depending on the electron concentration were obtained by Dhara
\emph{et al.}\cite{Dhara09} Very recently, on gated InAs nanowires
$l_{so}$ values between 50~nm and 150~nm depending on the nanowire
diameters were reported by Roulleau \emph{et al.}\cite{Roulleau10}
Thus our data obtained for short InAs nanowires showing strong
magneto-conductance fluctuations are in good agreement with the
experimental values previously reported for other types of InAs
nanowire systems. This demonstrates the reliability of our
approach.

For a planar surface two-dimensional electron gas in InAs the
Rashba parameter $\alpha_R$ was evaluated theoretically by Lamari.
\cite{Lamari03} For an electron concentration comparable to our
$\alpha_R \approx 1.1 \times 10^{-11}$~eVm was calculated.
Following the approach of Lamari we assessed $\alpha_R$ based on
the electric field $\mathcal{E}(r)$ averaged over all occupied
states $\psi_{nl}(r)$: $ \langle \mathcal{E} \rangle= \sum_{nl}
\langle
\psi_{nl}(r)|\mathcal{E}(r)|\psi_{nl}(r)\rangle$.\cite{Schaepers98b}
The calculated electric field is shown in Fig.~\ref{fig:2}(b). The
Rashba coupling parameter was estimated by using the following
expression:\cite{Engels97}
\begin{equation}
\alpha_R=\frac{E_p}{6m_0} \langle \mathcal{E}\rangle \left(
\frac{1}{E_g^2}-\frac{1}{(E_g+\Delta_{so})^2}\right) \label{eq:3}
\; ,
\end{equation}
with $E_p$ the interaction parameter for InAs, while $E_g$ and
$\Delta_{so}$ are the band gap and the valence band spin-orbit
splitting, respectively.\cite{Landolt02} Using Eq.~(\ref{eq:3}) we
obtained $\alpha_R=6.8 \times 10^{-12}$~eVm, resulting in a
spin-orbit scattering length $l_{so}=\hbar^2/(2m^*\alpha_R)$ of
170~nm. The fact that the value for $\alpha_R$ obtained in our
calculation is smaller than the ones experimentally obtained by
Schierholz \emph{et al.}\cite{Schierholz04} and the ones
calculated by Lamari \cite{Lamari03} might be due to the small
radius of the wire, since for small radii the Fermi level pinning
at the opposite side of the wire leads to an effective softening
of the potential profile. In order to check this assumption, we
performed a simulation for InAs wires with a sufficiently large
diameter so that the shape of the interface potential approaches
the one for a planar InAs surface. Here, we found that the values
of $\alpha_R$ obtained by Lamari\cite{Lamari03} are reproduced.
However, one should realize that the assumption of an infinitive
interface barrier is oversimplified.\cite{Winkler03} In fact, for
a more rigorous treatment the details of the surface potential
profile has to be included. Anyhow, since the surface of the InAs
is directly exposed to the environment, no reliable assumption can
be made at this stage. This is probably the reason for the larger
value of $l_{so}=170$~nm resulting from $\alpha_R$ obtained from
Eq.~(\ref{eq:3}), compared to the value of 70~nm extracted from
the comparison of the experimental weak-antilocalization curves
with the simulations based on the Green's function approach.

\section{Conclusions} \label{Sect:Con}

In summary, we determined the relevant low-temperature transport
parameters of short InAs nanowires by analyzing the fluctuating
magnetoconductance in detail. At the lowest temperature of 0.5~K a
phase-coherence length $l_\phi$ of at least 300~nm was extracted
from the correlation field $B_c$. Above 2~K a decrease of $l_\phi$
proportional to $T^{-0.35}$ was observed, which is in accordance
with theoretical predictions. Information on the spin-orbit
scattering length $l_{so}$ was gained by averaging the
magnetoconductance fluctuations over different gate voltages and
comparing these results to simulations based on a recursive
Green's function approach. We find that $l_{so}$ is in the order
of 70~nm indicating the presence of strong spin-orbit coupling.
Our investigations thus clearly prove the presence of spin-orbit
coupling in short InAs nanowires, which is an important
prerequisite for the realization of future spin electronic devices
based on semiconductor nanowires.


We gratefully acknowledge M. Governale (Victoria University of
Wellington, New Zealand) and A. Bringer (Institute of Solid State
Research, Forschungszentrum J\"ulich) for fruitful discussions.
This work was financial support by the French ANR (Quantamonde
project) and by DFG through FOR 912.

\end{document}